**Authors**

Hande Boyaci Selcuk*, Gabriella Reggiano*, Jacob Robson-Tull*, Lichirui Zhang*, João PGLM Rodrigues#

**Affiliation**

Schrödinger Inc, New York, NY, USA

*Authors contributed equally

# corresponding author: joao.rodrigues@schrodinger.com

**Title**

Towards better structural models from cryo-electron microscopy data with physics-based methods.

**Abstract**

Cryo-electron microscopy can now routinely deliver atomic resolution structures for a variety of biological systems. The relevance and value of these structures is directly related to their ability to help rationalize experimental observables, which in turn depends on the quality of model built into the density map. Coupling traditional model building tools with physics-based methods, such as docking, simulation, and modern force fields, has been shown to improve the quality of the resulting structures. Here, we survey the landscape of these hybrid approaches, highlighting their usefulness for medium- and low-resolution datasets, as well as for structures of small molecules, and make the argument that the community stands to benefit from their inclusion in model building and refinement workflows.

**Introduction**

Technological breakthroughs in the early 2010s – the so-called 'resolution revolution' – propelled cryogenic-electron microscopy (cryo-EM) into the mainstream of structural biology[1,2]. Driven by the introduction of direct electron detectors that drastically improved signal-to-noise ratio in image acquisition, and by the development of new image processing algorithms powered by graphical processing units, the median resolution of cryo-EM structures deposited in the wwPDB dropped from 9 Å in 2014, to slightly above 3 Å in 2024[3] (Fig. 1A). In addition, idiosyncrasies of cryo-EM, such as lower sample requirements, its ability to image molecules in (near-)native environments such as membranes, and the possibility of imaging multiple conformations in a single experiment, led to its application to a wide range of protein families, many of which had long remained recalcitrant to crystallization.

Among the protein families whose structural repertoires were enabled or expanded by cryo-EM, several belong to particularly relevant drug targets, such as G protein-coupled receptors (GPCRs), the protein tau, and different classes of ion channels[4,5]. GPCRs, the targets of more than a third of all approved drugs, have nearly 1,700 structures deposited in the wwPDB, 70% of which were determined by cryo-EM. More importantly, cryo-EM provided unprecedented access to active state receptors, bound to heterotrimeric G proteins, allowing for a broader understanding of GPCR biology and biased signaling. For protein tau, cryo-EM enabled atomic resolution structures of tau filaments and fibrils present in Alzheimer's disease, and consequently led to the rational design of small molecules that destabilize tau amyloid fibrils[6]. For $GABA_A$ receptors, the targets of benzodiazepines, barbiturates, and certain anesthetics, high-resolution cryo-EM structures provided key insights into receptor architecture, multimerization, and ligand binding and specificity for receptor subtypes[1]. In the case of the human ether-à-go-go-related potassium channel (hERG), a key anti-target in drug discovery whose inhibition leads to severe cardiotoxicity, several recent cryo-EM full-length structures are helping design new drugs with reduced side effects[7].

As a result, and unsurprisingly, the biotechnology and pharmaceutical industries have started to look to cryo-EM, by itself or alongside X-ray crystallography, as a source of high-resolution structural data for drug discovery. In a drug discovery setting, structures can reveal potential binding sites, illustrate conformational changes triggered by the binding of small molecule partners, and as such, inform the design of chemical matter that modulates target function[7,8]. The usefulness of a structure is, however, determined by its ability to rationalize experimental observables, which is, in turn, dependent on the quality of the structure and of the data it was modeled from. Therefore, robust tools for building and refining cryo-EM structures are critical, not only to success in a drug discovery setting, but also to the success of any studies relating structure to function.

**The importance of structure quality in cryo-EM**

As of 2025, nearly a third of cryo-EM structures (32.9%) were determined at a resolution equal to or under 3 Å, the threshold often cited as a prerequisite for structure-based drug design. At or below this resolution, modelers might be able to build amino acid side chains, ligands, and in some cases, even ions and water molecules. But just because they *can* do it does not mean they always *should*. Ambiguities and discontinuities in, or absences of, densities in cryo-EM maps lead to exercises in discretion by the modeler that can (and will) lead to subtle and seemingly innocuous mistakes that can have severe consequences in downstream applications[9].

Consider the structure and corresponding cryo-EM map of the human ACE2 receptor bound to the SARS-CoV-2 Spike protein, determined at 2.9 Å resolution (PDB 6M17; EMDB 30039)[10]. At first glance, the structure seems to be reasonably well-modeled into the density: both the receptor transmembrane and extracellular domains fit nicely in the map, while the viral protein domains are well-placed relative to the receptor (Fig. 1B, bottom). Upon closer inspection, there are several clashes between atoms of the receptor extracellular domains, as well as between sidechains of the receptor and viral protein. Moreover, the structure includes atoms for entire residues at the interface of both proteins, even in regions where there is little or no density to support modeling (Fig. 1B, top).

As one of the first structures of this complex to be published and made available to the community, it was used in several studies that tried to identify key structural determinants of affinity and specificity of the virus[11,12]. However, when compared to a crystal structure of the same complex (PDB 6M0J, 2.45 Å)[13], significant differences were observed in multiple regions of the interface, including in several key residues bridging the receptor and viral protein domains[11]. As a result, researchers working solely from the cryo-EM structure might have missed important interactions in their analyses, potentially leading them to wrong or incomplete conclusions. The poorly modeled interfaces would have also been detrimental for any virtual screening or free-energy perturbation methods, two commonly used methods in drug discovery that require accurate starting structures.

The purpose of this example is not to cast any cloud on a particular structure – the authors of the 6M17 structure acknowledge some of these issues themselves in their publication. Instead, we wish to highlight that structure *quality* is, generally, substantially more important than structure *completeness*. Whereas modelers and other experts can readily identify problematic regions in a cryo-EM structure, and even correct some of these issues, casual users will not, and instead rely on insufficient global metrics like resolution to gauge structure quality[14]. One particularly poignant example of this blind trust in resolution is its use as the sole metric to filter structures used to train structure prediction methods like AlphaFold. Moreover, since there is no active effort to automatically re-model cryo-EM structures with newer, more accurate methods, similar to PDB-REDO[15], any errors introduced in deposited models are likely to persist and possibly propagate.

Nonetheless, we believe that there are approaches that, if employed routinely, can be used to prevent – in the hands of modelers – or correct – in the hands of users – some of these errors in cryo-EM structures, and therefore improve the odds of success of most downstream applications.

**Building better models from cryo-EM density maps**

Most cryo-EM studies start with an analysis of tens of thousands of images of individual particles of the target molecule(s) of interest. A series of processing steps correct, sort, and align these images, leading to a 3D reconstruction that eventually becomes the density map onto which an initial model is built, either *de novo*, or through fitting an existing related (or predicted) structure. This initial model is then refined iteratively, against the density and a set of geometric restraints, until convergence.

In recent years, leveraging massive public datasets of sequences and structures, machine learning methods for structural biology overhauled the first steps of model building[16–18]. Structure prediction methods like AlphaFold trivialized the generation of highly accurate models of single- and multiple-domain proteins that can be fit into density maps and kickstart the modeling process. Alternatively, if the map is of enough quality, *de novo* approaches such as ModelAngelo[19] or DiffModeller[20], can build (entire) 3D structures of proteins and nucleic acids using only the density and sequence information.

Yet, despite being generally accurate, these methods are not without flaws. AlphaFold, for instance, has been shown to generate models with severe clashes, poor stereochemical geometry, and poor sidechain packing at protein interfaces and ligand binding pockets[17,21,22]. Structure refinement usually corrects these issues, particularly if the underlying density is strong and unambiguous. If the underlying density is poor, or absent, refinement is mostly reliant on the initial model and on the set of geometric restraints provided by the software, which can be insufficient to produce a high-quality structure by itself.

*Force field refinement substantially improves structure quality*

Geometric restraints are usually defined from high-resolution small molecule and macromolecule crystal data and include ideal bond lengths and angles, ideal torsions, definitions of planarity and chirality, and repulsive terms to prevent clashes between atoms[23]. These sets of restraints, very similar among different refinement programs, usually represent standard amino acids and nucleic acids, as well as common co-factors and other small molecules. For novel or rarer molecules, users need to generate bespoke geometric restraints, which is a challenging process often requiring specialized software and substantial technical and scientific expertise[24,25]. Nonetheless, the overall accuracy of these geometric restraints is limited as evidenced by the fact that lower resolution structures tend to have worse quality metrics, such as more clashes.

An alternative to using geometric restraints is to couple structure refinement with molecular mechanics force fields, which were developed to predict the structure and dynamics of macromolecules. Unlike early efforts, half a century ago[26], modern force fields make use of better and faster algorithms and are parameterized using large sets of *ab initio* quantum mechanical data[27,28]. They also offer a more complete depiction of nonbonded interactions, in contrast with geometrical restraints, such as electrostatics and both attractive and repulsive van der Waals forces. In addition, some force fields include treatment of solvent, which helps improve interactions between charged species. As a result, modern force fields are fairly accurate at reproducing experimental observables. The OPLS4 force field[29], for example, developed by our colleagues at Schrödinger, can predict the affinity of small molecule binding to cognate receptors to an accuracy approaching that of experiments[30]. Moreover, nearly every major force field – AMBER[31], CHARMM[32], OPLS[29], GROMOS[33] – makes available one or more tools that enable users to generate parameters for novel molecules in an automatic and self-consistent manner.

As a result of these advantages of force fields over standard geometric restraints, several groups have developed methods that combine specific force fields with structure refinement algorithms[34–38]. For instance, our team develops Phenix/OPLS[38], which pairs the structure refinement program Phenix with OPLS4 and the VSGB2.0 implicit solvent model that was parameterized for accurate structure prediction[39]. Phenix can also be used with the standard AMBER force field[37], or paired with Rosetta for fully automated model building[36]. Other approaches eschew standard refinement algorithms and instead integrate density cross-correlation (or similar) terms into structure prediction or molecular dynamics simulation software[40–43]. Finally, one last approach we believe noteworthy of inclusion, given its originality and popularity, is ISOLDE[44], which builds on OpenMM[45] and ChimeraX[46] to provide an interactive and physically realistic modeling environment.

In general, all these tools are able to improve the agreement between an initial model and the experimental density, while also improving several structure quality metrics, such as conformational strain, stereochemistry, and number of clashes (Fig. 2). On rigorous benchmarks strictly comparing the effects of replacing geometrical restraints by force fields, several independent reports show that the latter consistently improve

Molprobity scores, ClashScore, and Rama-Z scores at little to no cost of density cross-correlation, particularly for medium- and low-resolution maps[37,38]. Furthermore, our team has also shown that the use of force fields can help prevent overfitting to the density, as reported in a small dataset of 15 cryo-EM structures[38].

*Automated and robust small molecule placement with density-guided docking*

Another challenging aspect of building good quality models into cryo-EM density maps concerns small molecule ligands[47,48]. First, local map resolution is often worse in regions occupied by small molecules, in part because the image processing that precedes map reconstruction is biased by larger molecules, like proteins, but also because small molecules tend to be more flexible and have higher conformational heterogeneity. Even identifying *where* small molecules should be placed in the density can be a difficult task, as shown by reports of structures with waters modeled into ligand densities[49]. Then, whereas methods like AlphaFold are well-suited to predict the structure of proteins and nucleic acids, predicting small molecule binding remains an unsolved problem largely due to the sheer size and diversity of small molecule chemistry – the ChEMBL database[50] contains over 800,000 unique molecular scaffolds, for a total of 2.5 million entries, compared to 1,500 unique domains assigned to the 205,000 proteins entries deposited in the wwPDB. As a result, these prediction methods tend to produce small molecule models with bad geometries and/or in wrong binding poses or even pockets, particularly for 'unseen' molecules (or receptors) sufficiently different from those in their training sets[51].

On the other hand, the drug discovery community has dealt with small molecules for over three decades, and has an arsenal of computational methods that can be readily used to alleviate some of these issues. Unsurprisingly, in the last few years, several academic and industry teams, including our own, have developed approaches that target specifically the modelling of small molecule ligands into density maps. These approaches range from interactive tools, like the 'jiggle-fit' function in Coot[25], semi-automatic tools that fit ligands into densities and help kickstart the modeling process, like AFITT by OpenEye[52], to fully automated pipelines that combine fitting with structure refinement of both receptor and ligand, like EMERALD[53], ChemEM[34], GOLEM[35], and GemSpot[49]. The latter is a combination of several tools developed by our team at Schrödinger, namely GlideEM for ligand fitting and Phenix/OPLS for flexible refinement.

Barring nuances in scoring functions and optimization algorithms, ligand fitting is implemented similarly for all these approaches: there is an initial step that generates a set of ligand conformations, followed by a (more or less) exhaustive search of a pre-defined region that produces a small set of poses that simultaneously look chemically plausible and fit the density. To avoid misrepresenting another group's work, we will use GlideEM as a paradigmatic example to describe these steps in more detail, and refer the reader to the original publications of other methods for more details.

Starting from an input receptor structure, a cryo-EM density map, and a point in space defining the center of a region to dock the ligand into, GlideEM performs a cascade of increasingly stricter searches that samples millions of poses[54] and culminates in a set of (at most) 5 that have a combination of good fit to the density, good ligand stereochemistry, and good inter-molecular energies between ligand atoms and the surrounding environment. Fit to density is evaluated differently depending on the stage of the workflow: in earlier stages with tens of thousands of candidate poses, GlideEM uses an envelope-like metric that rewards overlap with density-rich regions of the map; later, during conformational optimization and energy minimization stages, performed only on tens or hundreds of poses, GlideEM switches to a more accurate, but slower, cross-correlation metric. This entire process takes less than 5 minutes on an average laptop.

Some of these methods focus entirely on the ligand and either keep the receptor rigid throughout the fitting process (GlideEM) or ignore the receptor altogether (AFITT). Others allow varying degrees of receptor flexibility, ranging from small side chain adjustments (GOLEM) to full flexibility of backbone and side chain atoms (EMERALD, ChemEM).

Lastly, although each of these tools was tested on largely independent datasets, owing to different dataset picking criteria and publication dates, the general consensus is that all of them are largely able to reproduce most of the deposited ligand poses to within 2 Å RMSD (excluding hydrogen atoms), with success rates depending primarily on the number of rotatable bonds of the ligand and map resolution. In our own experiments, a recent test on 426 cryo-EM structures of ligand-bound GPCRs with resolutions ranging from 2.3 Å to 4.5 Å showed that GlideEM finds a top-ranked solution within 1.5 Å and 2.5 Å RMSD of the deposited pose for 71% and 85% of the cases. If analyzing the top 5 poses returned by the software, the success rates increase to 83% and 92%, respectively (Fig. 2A). In addition, poses returned by GlideEM (as well as those of other tools) tend to have substantially fewer intra- and inter-molecular clashes compared to the deposited structures.

**Conclusion and Perspectives**

The 'resolution revolution' supercharged cryo-EM, enabling high-resolution structural biology of a wide and diverse range of new protein families. The ability to obtain atomic resolution structures of several previously uncharacterized, important drug targets is also leading to the use of cryo-EM structures in drug discovery programs. Here, structure quality is paramount[7,30] – a single misplaced sidechain can derail virtual screening campaigns or lead to poor results on free-energy perturbation simulations.

Given the heterogeneity in cryo-EM maps[55], even at high resolution there will be regions of ambiguous or sparse density, which force the modeler – or the software – to make assumptions (and errors)[9]. Despite efforts to create and promote metrics that evaluate the quality of a structure[47,56–58], in terms of its faithfulness to the density and of its stereochemistry, detecting such errors remains a non-trivial task.

Based on our own experience, both as developers and users of these tools, incorporating physics-based methods into structure refinement is a simple and effective approach to improving the overall geometric and stereochemical quality of models built from cryo-EM maps, particularly those containing small molecule ligands. Several existing implementations, many of which are already available on SBGrid[59], offer a range of semi and fully automated solutions that produce poses on par with, or better than, human experts in a few minutes or hours.

We believe these approaches will become increasingly relevant, particularly as advances in cryo-EM and machine learning multiply our ability to produce high-resolution data for a variety of biological systems and conditions[18,60,61].

**References**

1 de Oliveira TM, van Beek L, Shilliday F, Debreczeni JÉ & Phillips C (2021) Cryo-EM: The Resolution Revolution and Drug Discovery. *SLAS Discov* **26**, 17–31.
2 Bai X, McMullan G & Scheres SHW (2015) How cryo-EM is revolutionizing structural biology. *Trends Biochem Sci* **40**, 49–57.
3 Burley SK, Berman HM, Chiu W, Dai W, Flatt JW, Hudson BP, Kaelber JT, Khare SD, Kulczyk AW, Lawson CL, Pintilie GD, Sali A, Vallat B, Westbrook JD, Young JY & Zardecki C (2022) Electron microscopy holdings of the Protein Data Bank: the impact of the resolution revolution, new validation tools, and implications for the future. *Biophys Rev* **14**, 1281–1301.
4 Robertson MJ, Meyerowitz JG & Skiniotis G (2022) Drug discovery in the era of cryoEM. *Trends Biochem Sci* **47**, 124–135.
5 Renaud J-P, Chari A, Ciferri C, Liu W, Rémigy H-W, Stark H & Wiesmann C (2018) Cryo-EM in drug discovery: achievements, limitations and prospects. *Nat Rev Drug Discov* **17**, 471–492.
6 Seidler PM, Murray KA, Boyer DR, Ge P, Sawaya MR, Hu CJ, Cheng X, Abskharon R, Pan H, DeTure MA, Williams CK, Dickson DW, Vinters HV & Eisenberg DS (2022) Structure-based discovery of small molecules that disaggregate Alzheimer's disease tissue derived tau fibrils in vitro. *Nat Commun* **13**, 5451.
7 Miller EB, Hwang H, Shelley M, Placzek A, Rodrigues JPGLM, Suto RK, Wang L, Akinsanya K & Abel R (2024) Enabling structure-based drug discovery utilizing predicted models. *Cell* **187**, 521–525.
8 Wei H & McCammon JA (2024) Structure and dynamics in drug discovery. *Npj Drug Discov* **1**, 1–8.
9 Gao Y, Thorn V & Thorn A (2023) Errors in structural biology are not the exception. *Acta Crystallogr Sect Struct Biol* **79**, 206–211.
10 Yan R, Zhang Y, Li Y, Xia L, Guo Y & Zhou Q (2020) Structural basis for the recognition of SARS-CoV-2 by full-length human ACE2. *Science* **367**, 1444–1448.
11 Rodrigues JPGLM, Barrera-Vilarmau S, M. C. Teixeira J, Sorokina M, Seckel E, Kastritis PL & Levitt M (2020) Insights on cross-species transmission of SARS-CoV-2 from structural modeling. *PLOS Comput Biol* **16**, e1008449.
12 Sorokina M, Belapure J, Tüting C, Paschke R, Papasotiriou I, Rodrigues JPGLM & Kastritis PL (2022) An Electrostatically-steered Conformational Selection Mechanism Promotes SARS-CoV-2 Spike Protein Variation. *J Mol Biol* **434**, 167637.
13 J L, J G, J Y, S S, H Z, S F, Q Z, X S, Q W, L Z & X W (2020) Structure of the SARS-CoV-2 spike receptor-binding domain bound to the ACE2 receptor. *Nature* **581**.
14 Lander GC (2024) Single particle cryo-EM map and model validation: It's not crystal clear. *Curr Opin Struct Biol* **89**, 102918.
15 Joosten RP, Joosten K, Cohen SX, Vriend G & Perrakis A (2011) Automatic rebuilding and optimization of crystallographic structures in the Protein Data Bank. *Bioinformatics* **27**, 3392–3398.
16 Terashi G, Wang X & Kihara D (2023) Protein model refinement for cryo-EM maps using AlphaFold2 and the DAQ score. *Acta Crystallogr Sect Struct Biol* **79**, 10–21.
17 Terwilliger TC, Liebschner D, Croll TI, Williams CJ, McCoy AJ, Poon BK, Afonine PV, Oeffner RD, Richardson JS, Read RJ & Adams PD (2024) AlphaFold predictions are valuable hypotheses and accelerate but do not replace experimental structure determination. *Nat Methods* **21**, 110–116.
18 Tüting C, Schmidt L, Skalidis I, Sinz A & Kastritis PL (2023) Enabling cryo-EM density interpretation from yeast native cell extracts by proteomics data and AlphaFold structures. *PROTEOMICS* **23**, 2200096.
19 Jamali K, Käll L, Zhang R, Brown A, Kimanius D & Scheres SHW (2024) Automated model building and protein identification in cryo-EM maps. *Nature* **628**, 450–457.
20 Wang X, Zhu H, Terashi G, Taluja M & Kihara D (2024) DiffModeler: large macromolecular structure modeling for cryo-EM maps using a diffusion model. *Nat Methods* **21**, 2307–2317.
21 Coskun D, Lihan M, Rodrigues J, Vass M, Robinson D, Friesner R & Miller E (2023) Using AlphaFold and Experimental Structures for the Prediction of the Structure and Binding Affinities of GPCR Complexes via In-duced Fit Docking and Free Energy Perturbation. .
22 Träger TK, Tüting C & Kastritis PL (2024) The human touch: Utilizing AlphaFold 3 to analyze structures of endogenous metabolons. *Structure* **32**, 1555–1562.
23 Moriarty NW, Grosse-Kunstleve RW & Adams PD (2009) electronic Ligand Builder and Optimization Workbench (eLBOW): a tool for ligand coordinate and restraint generation. *Acta Crystallogr D Biol Crystallogr* **65**, 1074–1080.
24 Liebschner D, Moriarty NW, Poon BK & Adams PD (2023) In situ ligand restraints from quantum-mechanical methods. *Acta Crystallogr Sect Struct Biol* **79**, 100–110.
25 Debreczeni JÉ & Emsley P (2012) Handling ligands with Coot. *Acta Crystallogr D Biol Crystallogr* **68**, 425–430.
26 Levitt M (1974) Energy refinement of hen egg-white lysozyme. *J Mol Biol* **82**, 393–420.

27 Nerenberg PS & Head-Gordon T (2018) New developments in force fields for biomolecular simulations. *Curr Opin Struct Biol* **49**, 129–138.
28 van Gunsteren WF, Dolenc J & Mark AE (2008) Molecular simulation as an aid to experimentalists. *Curr Opin Struct Biol* **18**, 149–153.
29 Lu C, Wu C, Ghoreishi D, Chen W, Wang L, Damm W, Ross GA, Dahlgren MK, Russell E, Von Bargen CD, Abel R, Friesner RA & Harder ED (2021) OPLS4: Improving Force Field Accuracy on Challenging Regimes of Chemical Space. *J Chem Theory Comput* **17**, 4291–4300.
30 Ross GA, Lu C, Scarabelli G, Albanese SK, Houang E, Abel R, Harder ED & Wang L (2023) The maximal and current accuracy of rigorous protein-ligand binding free energy calculations. *Commun Chem* **6**, 1–12.
31 Maier JA, Martinez C, Kasavajhala K, Wickstrom L, Hauser KE & Simmerling C (2015) ff14SB: Improving the Accuracy of Protein Side Chain and Backbone Parameters from ff99SB. *J Chem Theory Comput* **11**, 3696–3713.
32 Huang J, Rauscher S, Nawrocki G, Ran T, Feig M, de Groot BL, Grubmüller H & MacKerell AD (2017) CHARMM36m: An Improved Force Field for Folded and Intrinsically Disordered Proteins. *Nat Methods* **14**, 71–73.
33 Oostenbrink C, Villa A, Mark AE & van Gunsteren WF (2004) A biomolecular force field based on the free enthalpy of hydration and solvation: the GROMOS force-field parameter sets 53A5 and 53A6. *J Comput Chem* **25**, 1656–1676.
34 Sweeney A, Mulvaney T, Maiorca M & Topf M (2024) ChemEM: Flexible Docking of Small Molecules in Cryo-EM Structures. *J Med Chem* **67**, 199–212.
35 Zhao Z & Tajkhorshid E (2024) GOLEM: Automated and Robust Cryo-EM-Guided Ligand Docking with Explicit Water Molecules. *J Chem Inf Model* **64**, 5680–5690.
36 Wang RY-R, Song Y, Barad BA, Cheng Y, Fraser JS & DiMaio F (2016) Automated structure refinement of macromolecular assemblies from cryo-EM maps using Rosetta. *eLife* **5**, e17219.
37 Moriarty NW, Janowski PA, Swails JM, Nguyen H, Richardson JS, Case DA & Adams PD (2020) Improved chemistry restraints for crystallographic refinement by integrating the Amber force field into Phenix. *Acta Crystallogr Sect Struct Biol* **76**, 51–62.
38 van Zundert GCP, Moriarty NW, Sobolev OV, Adams PD & Borrelli KW (2021) Macromolecular refinement of X-ray and cryoelectron microscopy structures with Phenix/OPLS3e for improved structure and ligand quality. *Structure* **29**, 913-921.e4.
39 Li J, Abel R, Zhu K, Cao Y, Zhao S & Friesner RA (2011) The VSGB 2.0 model: a next generation energy model for high resolution protein structure modeling. *Proteins* **79**, 2794–2812.
40 Blau C, Yvonnesdotter L & Lindahl E (2022) Gentle and fast all-atom model refinement to cryo-EM densities via Bayes' approach. .
41 Shugaeva T, Howard RJ, Haloi N & Lindahl E (2025) Modeling cryo-EM structures in alternative states with generative AI and density-guided simulations. .
42 Yvonnesdotter L, Rovšnik U, Blau C, Lycksell M, Howard RJ & Lindahl E (2023) Automated simulation-based membrane protein refinement into cryo-EM data. *Biophys J* **122**, 2773–2781.
43 Trabuco LG, Villa E, Mitra K, Frank J & Schulten K (2008) Flexible fitting of atomic structures into electron microscopy maps using molecular dynamics. *Struct Lond Engl 1993* **16**, 673–683.
44 Croll TI (2018) ISOLDE: a physically realistic environment for model building into low-resolution electron-density maps. *Acta Crystallogr Sect Struct Biol* **74**, 519–530.
45 Eastman P, Galvelis R, Peláez RP, Abreu CRA, Farr SE, Gallicchio E, Gorenko A, Henry MM, Hu F, Huang J, Krämer A, Michel J, Mitchell JA, Pande VS, Rodrigues JP, Rodriguez-Guerra J, Simmonett AC, Singh S, Swails J, Turner P, Wang Y, Zhang I,

Chodera JD, De Fabritiis G & Markland TE (2024) OpenMM 8: Molecular Dynamics Simulation with Machine Learning Potentials. *J Phys Chem B* **128**, 109–116.

46 Pettersen EF, Goddard TD, Huang CC, Meng EC, Couch GS, Croll TI, Morris JH & Ferrin TE (2021) UCSF ChimeraX: Structure visualization for researchers, educators, and developers. *Protein Sci Publ Protein Soc* **30**, 70–82.

47 Lawson CL, Kryshtafovych A, Pintilie GD, Burley SK, Černý J, Chen VB, Emsley P, Gobbi A, Joachimiak A, Noreng S, Prisant MG, Read RJ, Richardson JS, Rohou AL, Schneider B, Sellers BD, Shao C, Sourial E, Williams CI, Williams CJ, Yang Y, Abbaraju V, Afonine PV, Baker ML, Bond PS, Blundell TL, Burnley T, Campbell A, Cao R, Cheng J, Chojnowski G, Cowtan KD, DiMaio F, Esmaeeli R, Giri N, Grubmüller H, Hoh SW, Hou J, Hryc CF, Hunte C, Igaev M, Joseph AP, Kao W-C, Kihara D, Kumar D, Lang L, Lin S, Subramaniya SRMV, Mittal S, Mondal A, Moriarty NW, Muenks A, Murshudov GN, Nicholls RA, Olek M, Palmer CM, Perez A, Pohjolainen E, Pothula KR, Rowley CN, Sarkar D, Schäfer LU, Schlicksup CJ, Schröder GF, Shekhar M, Si D, Singharoy A, Sobolev OV, Terashi G, Vaiana AC, Vedithi SC, Verburgt J, Wang X, Warshamanage R, Winn MD, Weyand S, Yamashita K, Zhao M, Schmid MF, Berman HM & Chiu W (2024) Outcomes of the EMDataResource Cryo-EM Ligand Modeling Challenge. *Nat Methods* **21**, 1340–1348.

48 Malhotra S, Träger S, Dal Peraro M & Topf M (2019) Modelling structures in cryo-EM maps. *Curr Opin Struct Biol* **58**, 105–114.

49 Robertson MJ, van Zundert GCP, Borrelli K & Skiniotis G (2020) GemSpot: A Pipeline for Robust Modeling of Ligands into Cryo-EM Maps. *Structure* **28**, 707-716.e3.

50 Gaulton A, Bellis LJ, Bento AP, Chambers J, Davies M, Hersey A, Light Y, McGlinchey S, Michalovich D, Al-Lazikani B & Overington JP (2012) ChEMBL: a large-scale bioactivity database for drug discovery. *Nucleic Acids Res* **40**, D1100–D1107.

51 Jain AN, Cleves AE & Walters WP (2024) Deep-Learning Based Docking Methods: Fair Comparisons to Conventional Docking Workflows. .

52 Wlodek S, Skillman AG & Nicholls A (2006) Automated ligand placement and refinement with a combined force field and shape potential. *Acta Crystallogr D Biol Crystallogr* **62**, 741–749.

53 Muenks A, Zepeda S, Zhou G, Veesler D & DiMaio F (2023) Automatic and accurate ligand structure determination guided by cryo-electron microscopy maps. *Nat Commun* **14**, 1164.

54 Friesner RA, Banks JL, Murphy RB, Halgren TA, Klicic JJ, Mainz DT, Repasky MP, Knoll EH, Shelley M, Perry JK, Shaw DE, Francis P & Shenkin PS (2004) Glide: A New Approach for Rapid, Accurate Docking and Scoring. 1. Method and Assessment of Docking Accuracy. *J Med Chem* **47**, 1739–1749.

55 Dubach VRA & Guskov A (2020) The Resolution in X-ray Crystallography and Single-Particle Cryogenic Electron Microscopy. *Crystals* **10**, 580.

56 Reggiano G, Lugmayr W, Farrell D, Marlovits TC & DiMaio F (2023) Residue-level error detection in cryo-electron microscopy models. *Struct Lond Engl 1993* **31**, 860.

57 Lawson CL, Kryshtafovych A, Adams PD, Afonine PV, Baker ML, Barad BA, Bond P, Burnley T, Cao R, Cheng J, Chojnowski G, Cowtan K, Dill KA, DiMaio F, Farrell DP, Fraser JS, Herzik MA, Hoh SW, Hou J, Hung L-W, Igaev M, Joseph AP, Kihara D, Kumar D, Mittal S, Monastyrskyy B, Olek M, Palmer CM, Patwardhan A, Perez A, Pfab J, Pintilie GD, Richardson JS, Rosenthal PB, Sarkar D, Schäfer LU, Schmid MF, Schröder GF, Shekhar M, Si D, Singharoy A, Terashi G, Terwilliger TC, Vaiana A, Wang L, Wang Z, Wankowicz SA, Williams CJ, Winn M, Wu T, Yu X, Zhang K, Berman HM & Chiu W (2021) Cryo-EM model validation recommendations based on outcomes of the 2019 EMDataResource challenge. *Nat Methods* **18**, 156–164.

58 Pintilie G, Shao C, Wang Z, Hudson BP, Flatt JW, Schmid MF, Morris K, Burley SK & Chiu W (2025) Q-score as a reliability measure for protein, nucleic acid, and small molecule atomic coordinate models derived from 3DEM density maps. *bioRxiv*, 2025.01.14.633006.
59 Morin A, Eisenbraun B, Key J, Sanschagrin PC, Timony MA, Ottaviano M & Sliz P (2013) Collaboration gets the most out of software. *eLife* **2**, e01456.
60 Kreis K, Dockhorn T, Li Z & Zhong E (2022) Latent Space Diffusion Models of Cryo-EM Structures. .
61 Raghu R, Levy A, Wetzstein G & Zhong ED (2025) Multiscale guidance of AlphaFold3 with heterogeneous cryo-EM data. .

## Figures

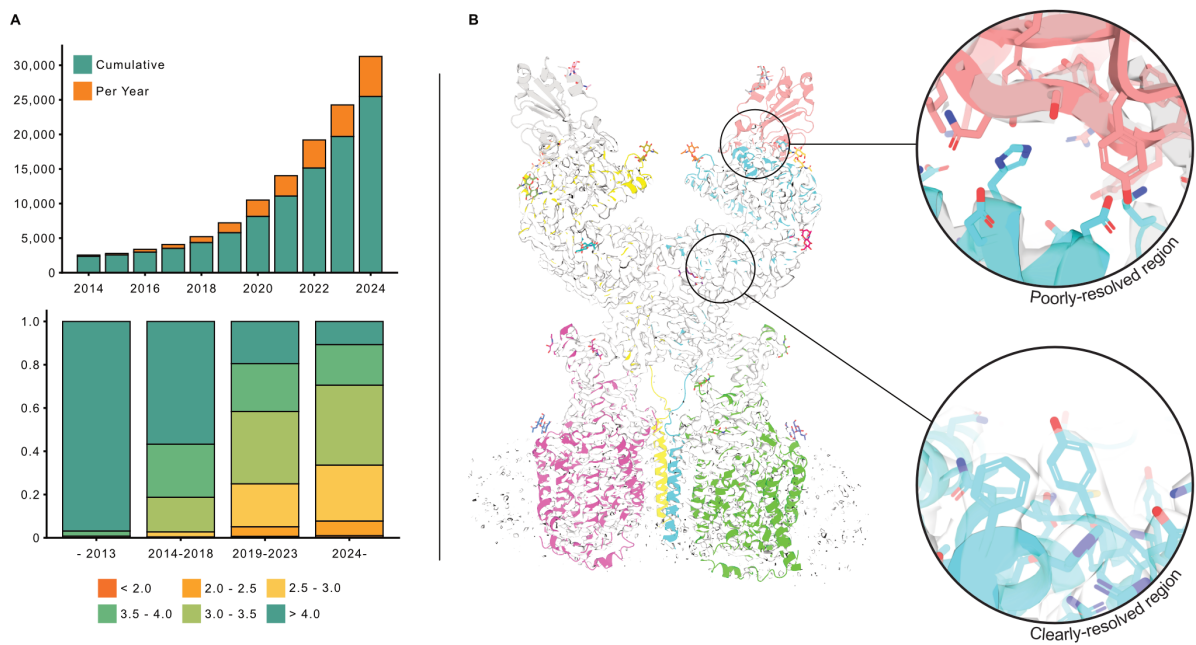

**Figure 1.** The 'resolution revolution', started ca. 2014, led to the establishment of cryo-EM as a mainstream method in structural biology. The number of structures determined by cryo-EM that were deposited in the wwPDB has been growing steadily (A, top), while the nominal resolution of depositions has been improving (A, bottom). As of 2025, the majority of cryo-EM structures was determined at or below 3 Å. However, as resolution is not constant across the map (B), even high-resolution structures (e.g. PDB 6M17, 2.9 Å, shown at author recommended isocontour) can have regions of poor density that warrant special care when modeling into.

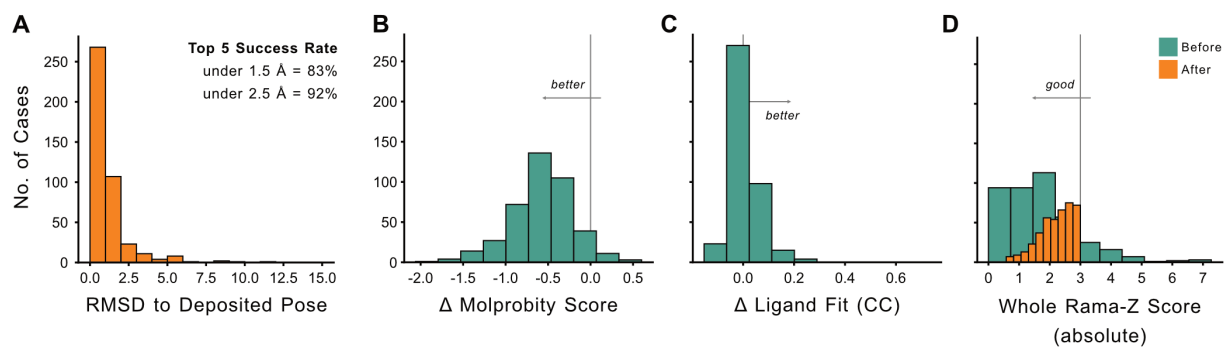

**Figure 2.** Physics-based methods for ligand placement (A) and structure refinement (B-D) help automate model building and improve stereochemical quality for cryo-EM structures. Density-guided ligand docking with GlideEM (A) can place ligands to within 2.5 Å of deposited poses for 92% of the cases in a dataset of 426 GPCR cryo-EM structures. Physics-based refinement with Phenix/OPLS improves the quality of deposited structures in terms of Molprobity Score (B) and Rama-Z score (D), at little to no expense of density cross-correlation (C).

## Acknowledgements

We would like to thank our colleagues Robert Abel, Alfie-Louise Brownless, Zachary Johnson, Michael Di Matia, and Edward Miller for their comments and suggestions.